# Novel recoil nuclei detectors to qualify the AMANDE facility as a Standard for mono-energetic neutron fields

A. Allaoua[a], O. Guillaudin[c], S. Higueret[b], D. Husson[b], L. Lebreton[a], F. Mayet[c], M. Nourreddine[b], D. Santos[c], A. Trichet[c].

[a] *Laboratory for neutron Metrology and Neutron dosimetry, Institute for Radiological Protection and Nuclear Safety, 13115 Saint Paul Lez Durance, France*

[b] *Institut Pluridisciplinaire Hubert Curien, UMR 7178, 23 rue du Loess, BP28, Strasbourg Cedex 2, France*

[c] *Laboratoire de Physique Subatomique et de Cosmologie, Université Joseph Fourier Grenoble 1, CNRS/IN2P3, Institut National Polytechnique de Grenoble, 53, avenue des Martyrs, 38026 Grenoble cedex, France*

**Abstract**

The AMANDE facility at IRSN-Cadarache produces mono-energetic neutron fields from 2 keV to 20 MeV with metrological quality. To be considered as a standard facility, characteristics of neutron field i.e fluence distribution must be well known by a device using absolute measurements. The development of new detector systems allowing a direct measurement of neutron energy and fluence has started in 2006. Using the proton recoil telescope principle with the goal of increase the efficiency, two systems with full localization are studied. A proton recoil telescope using CMOS sensor (CMOS-RPT) is studied for measurements at high energies and the helium 4 gaseous μ-time projection chamber (μ-TPC $^4$He) will be dedicated to the lowest energies. Simulations of the two systems were performed with the transport Monte Carlo code MCNPX, to choose the components and the geometry, to optimize the efficiency and detection limits of both devices or to estimate performances expected. First preliminary measurements realised in 2008 demonstrated the proof of principle of these novel detectors for neutron metrology.

## 1. Introduction

One of the main LMDN activities consists in developing and operating neutron facilities. One of them, AMANDE, is a 2 MV tandetron accelerator which produces standard mono energetic neutron fields from 2keV to 20 MeV (Gressier et al., 2004). The objective of AMANDE is to study the response of neutron dosimeters or detection systems as a function of incident energy. To achieve the level of a standard facility, absolute neutron fluence measurements are required, instead of methods using detector systems with simulated response function and unfolding methods. For this purpose, the development of new detectors acting as Recoil Nuclear Telescopes (RNT) has been started.

The RNT principle is based on the detection of the recoil nucleus emitted by the elastic scattering of the neutron on a thin converter foil. From the simple kinematics, the energy of the recoil nuclei ($E_r$) is related to the incident neutron energy ($E_n$) by the relationship: $E_n = (1+A)^2 \cdot E_r /(4 \cdot A \cdot \cos^2\theta)$ where $\theta$ is the angle of the recoil nuclei with respect to the incident neutron direction. The simultaneous measurement of the nucleus energy and recoil angle leads to the initial neutron energy. For some existing Recoil Proton Telescope (RPT) used in this energy range (Siebert et al., 1985), the recoil angle is fixed and the energy of the scattered proton is measured in a narrow solid angle generated by a small area detector. The precision on fluence is of 3 to 5%, depending on the neutron energy. However, due to the small acceptance angle, the efficiency is quite low (about 0.001%).

To improve this method, it is necessary is to enhance the efficiency by increasing the detection solid angle. The main idea is to use localization detector systems in order to get full spatial information on the nucleus scattering angle.
Because of a large energy range of AMANDE neutron fields, two systems are needed.

## 2. The CMOS-RPT principle

The detection energy range of the CMOS-RPT should be 6 MeV to 20 MeV. This challenge is to be reached with CMOS pixel systems, which are presently available as very thin detectors. The device is made up of a polyethylene neutron/proton converter foil, three CMOS active pixel sensor plates which give 3D localization at the micron level, followed by a thick Si(Li) diode to measure the proton energy (figure 1). The CMOS chips (Turchetta et al., 2001) are of the "MIMOSA" type (Minimum Ionising particles MOS Active pixel sensor) developed in Strasbourg. The large area chips are made up of two independent matrices of 320x320 pix². With a pixel pitch of 30x30 μm², the area of the two matrices is almost 2cm². The sensitive region of these chips consists of 15μm deep high crystalline quality epitaxial silicon layer. This region is sandwiched between a 5µm SiO$_2$ layer in the upper side and a non-epi silicon substrate. After post-processing of the sensors, the substrate is thinned down to 30µm. The efficiency of these sensors to few MeV protons is 100% since they have been optimised for minimum ionising particles (e.g. GeV protons). Of special interest for neutron detection, the thinness of the epitaxial layer makes them virtually transparent to gamma-rays. In this prototype, the Si(Li) is of only 1cm² and only one matrix per CMOS sensor is used.

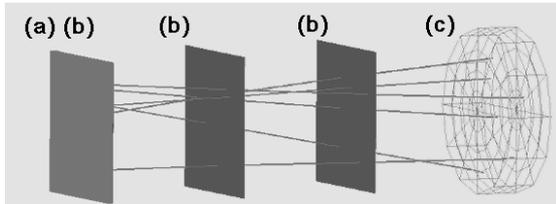

Fig 1. 3D visualization of the CMOS-RPT. (a) is the converter, (b) the CMOS planes and (c) the Si(Li) diode. The MCNPX simulation shows some tracks of protons hitting all detectors.

## 3. The CMOS-RPT calculation works

Most of the calculation made with MCNPX 2.6 Monte Carlo code (http://mcnpx.lanl.gov). The first study dealt with the estimation of the converter thickness depending on the neutron energy to optimize the efficiency and the accuracy. The thickness must be chosen in order to maximize the number of secondary protons and get acceptable the energy loss in the converter. For example, for a 15 MeV incident neutron energy, the thickness must be less than 360 µm if the relative proton energy loss at 0° is chosen be lower than 5% and 750 µm if the energy loss should be lower than 10%.

A second critical point is the distance between each CMOS planes, as the overall efficiency is strongly dependant on the acceptance angle. A distance of 1 cm can probably be achieved within the mechanical constraints: with this geometry, the maximum proton angle is 21° and for 15 MeV energy neutrons, the efficiency is calculated to be $7.10^{-5}$ for a 100 µm converter thickness.

The studies of the detection range and the energy resolution were also performed. The proton energy $E_P$ is the sum of the energy loss in the converter $E_{P1}$, the energy loss in the 3 CMOS planes and air layers $E_{P2}$ and the detected energy in the diode $E_{P3}$. Both energies $E_{P1}$ and $E_{P2}$ have been estimated by a C++ code using NIST Pstar data. The relative uncertainty associated to this calculated energy ($\Delta E_{calculated}/E_{calculated} > 10\%$) is higher than the detected one, for which an excellent resolution is obtained with high quality diodes ($\Delta E_{P3}/E_{P3} < 1\%$). The non-measured energy must be as small as possible to minimise the total uncertainty. As shown in figure 2, the ratio between the calculated energy and the total energy rises sharply when the proton energy decreases. Below 6MeV proton energy, the calculated part represents more than 50% of the total. For this reason, the working region is limited to above 6 MeV.

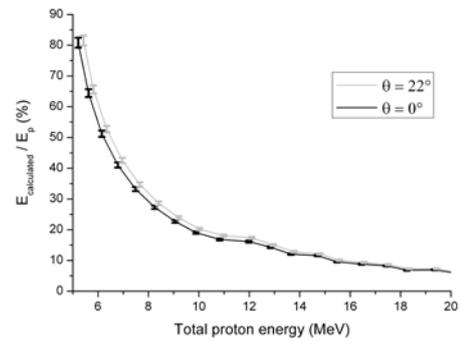

Fig 2. Ratio between the calculated energy and the total proton energy as a function of the total proton energy (for protons at minimum and maximum angle, with a 300 µm converter)

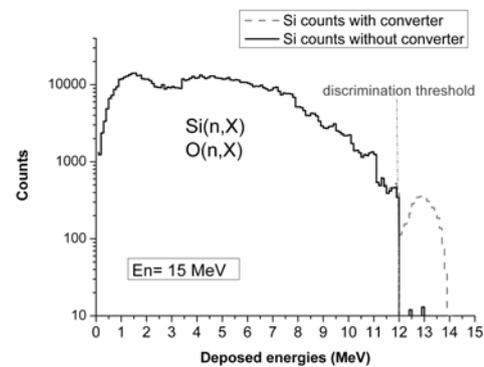

Fig 3. Deposed energy distribution of the charged particles in the diode with and without the converter (neutron energy: 15 MeV).

Discrimination between the proton coming from elastic scattering of neutrons in the converter and other secondary particles (expected protons from elastic scattering) is feasible by two ways. A first criterion is given by complete tracks, as the protons have to cross the three CMOS planes to be identified as good events. To evaluate the possibility of energy discrimination, some calculation using MCNPX code were done to evaluate the total charged particle counts as a function of the deposed energy in the Si(Li) diode. Calculation were done for two configuration, first with the complete telescope (the converter, the three CMOS planes and the Si(Li) diode), the second without the converter plane. The main result (figure 3) is that proton from scattered from the converter have the highest deposed energy in the Si(Li) diode. A threshold depending on the neutron energy can be chosen to select only the true recoil protons.

At low energies, the angular resolution can be strongly affected by multiple scattering, and this problem also was studied by simulation. Simulations of the counts per pixel from a narrow proton beam crossing the second and the third CMOS plane were done. A gaussian fit leads to a standard deviation of about 10 mrad for 20 MeV protons. This figure increases at lower energies but remains sustainable.

## 4. Experiment with the EUDET telescope

During the construction of first CMOS prototype, a first experiment was performed with an existing device. The EUDET telescope is a high resolution beam pixel system, developed within the EUDET collaboration

(http://www.eudet.org) as a beam test tool for new detectors. In this implementation, the telescope consisted in four CMOS planes with Mimosa 18 sensors (512 x 512 pixels with 10μm pitch and 60μm thinness). No diode was used here.

The main results of these tests, published elsewhere (Husson et al., 2008), are an excellent acceptance and efficiency, a detection capability down to 5 MeV neutrons and a multiple scattering below 80 mrad at all energies. Figure 4 shows an example of an experimental proton track and the associated angle as detected in EUDET.

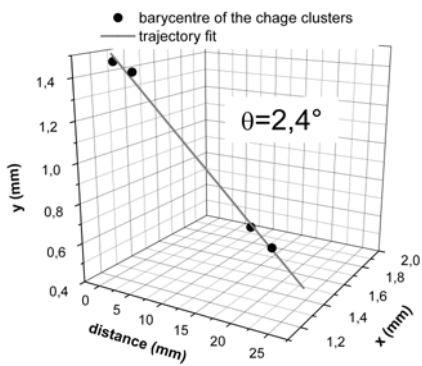

Fig 4. A measured proton track proton track and the associated angle detected in EUDET coming from converter

## 5. The μ-TPC $^4$He principle

For the low neutron energies produced by AMANDE facility, a bibliographic study leaded LMDN to choose a μ-TPC spectrometer. The detector of LPSC Grenoble (figure 5) is originally devoted to the detection of dark matter. Its volume is 10x10x20cm$^3$. For neutron detection, the set-up is adapted with a gaseous mixer of 95% $^4$He and 5% $C_4H_{10}$ and acts also as a p/n and α/n converter. Alpha particles and protons ionise gas, electric charges are then created and electrons drift from the drift chamber to the amplification chamber, where the avalanche takes place. The charges are picked up by the pixels anode plane (with a 250 μm pitch) and the 2D track is obtained. The third Dimension is provided by the drift times of charges. Thereby, the angle of the recoil is obtained. The sum of collected charges provides information about the amount of the ionization produced by the recoil. The ionization quenching factor (IQF) is defined as the fraction of energy released by recoil in a medium through ionization compared with its total kinetic energy. $^4$He IQF has recently been measured down to very low energies (1 keV) (Santos et al., 2008). Therefore, with the knowledge of the $^4$He IQF and the amount of ionization, the total energy of the recoil nucleus can be obtained. The detection energy range is between 2 keV and 5 MeV, depending on the gas pressure.

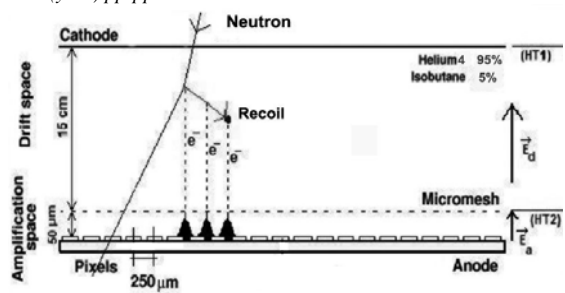

Fig 5. Sketch of the μ-TPC $^4$He

## 6. The μ-TPC $^4$He calculations

Before performing MCNPX simulations, a validation test of the recoil proton and alpha transport with this code was realised. The simulated angular distribution and the energy distribution of both recoil nuclei are in agreement with the theory. However, in the MCNPX version 2.6c, the simulated range of recoil nuclei in the gas, was in disagreement, below 100 keV, with NIST data. An update of the 2.6f version, taking into account range correction (new patch) gives a much better agreement.

TRIM (www.srim.org) calculations were done for two neutron energies. The 144 keV is an ISO-recommended energy available at AMANDE facility and 8.2 keV is the lowest energy obtained with a satisfying precision in this facility. The goal of these simulations is to calculate, for different pressures, the total track length inside the detection volume and the best angular determination (1 mm track with negligible multiple scattering). For 144 keV, the more adapted pressure is 1 bar and 350 mbar for 8.2 keV. Moreover, the efficiencies were estimated for both energies: $3.59 \cdot 10^{-4}$ with alpha particles and $2.48 \cdot 10^{-3}$ with protons at 144 keV and $2.28 \cdot 10^{-4}$ (α) and $3.26 \cdot 10^{-3}$ (p) at 8.2 keV. The good results for simulated efficiency come more from the recoil proton (from 5% isobutene) than from the recoil alpha.

## 7. First experimental results

The goal of the first measurement campaign was the validation of low energy neutron detection without pixellized anode. Measurements were performed with neutrons about 105 keV (mean energy). Several tests were performed (electrical behaviour of the detector, test of different kinds of quencher, electromagnetic noise effects, photons sensitivity) and leaded to optimise the system. The figure 7 shows the experimental ionisation energy distribution of the recoil nuclei. The proton and alpha recoils "end point" are easily separated: about 123 keV for protons and 73 keV for the alpha particles.

The figure 8 shows the MCNPX simulated total energy distribution of the recoil nuclei from the neutrons incident distribution (measured by SP2 spectrometer). The recoil nuclei "end points" are easy to identify: about 133 keV for protons and 80 keV for the alpha particles. The energy shifts are due to the $^4$He IQF.

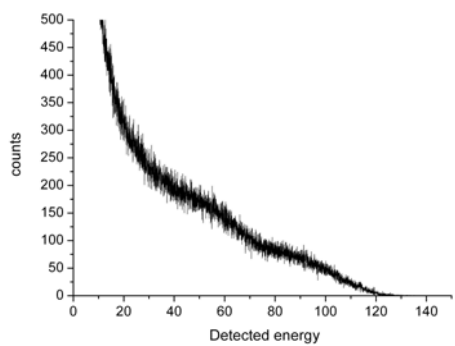

Fig 7. Experimental ionisation energy distribution of the recoil nuclei from neutrons (mean neutron energy: 105 keV)

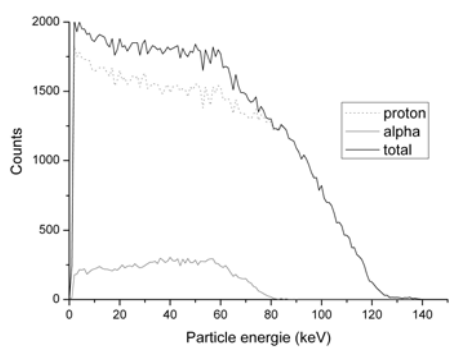

Fig 8. MCNPX simulation of the energy distribution for recoil proton, for recoil alpha nuclei and for both recoil (mean neutron energy: 105 keV)

## 8. Achievements and perspectives

The simulations of both detectors were performed to optimise the efficiency and the detection limits. Theoretical efficiencies obtained are better than for a classical RPT. But the study of the neutron energy uncertainty must be ended up to compare with the classical zero-degree RPT. The first measurement campaign was encouraging for our own CMOS-RPT and the next experimental tests are planned for 2009 with the first prototype. The estimation of the energetic distribution of the fluence will be possible.

For the µ-TPC, simulations with GEANT4 or COMSOL software will be performed to study the influence of the electric and magnetic fields. With this detector, the first measurement campaign was a satisfying test for the neutron detection. A second measurement campaign was performed using a pixellized anode. In spite of a limited number of active pixels, some tracks were observed. Analysis is in progress. New measurements on the AMANDE facility are scheduled next year.

**Acknowledgments**

This work is supported by the French national laboratory of Metrology (LNE). We would like to thank V. Gressier, A. Martin and M. Pepino, all of them belonging to LMDN, for their unmitigated help during experiments on AMANDE facility. We are greatly indebted to the CERN/EUDET team: W.Dulinski for hardware, D.Haas for the DAQ system and E.Corrin for his software expertise.